\documentclass[twocolumn,showpacs,preprintnumbers,amsmath,amssymb]{revtex4}

\usepackage{graphicx}
\usepackage{dcolumn}
\usepackage{bm}

\begin{document}


\title{Linear scaling between momentum- and spin scattering in graphene}

\author{C. J\'{o}zsa$^1$}
\author{T. Maassen$^1$, M. Popinciuc$^2$, P.J. Zomer$^1$, A. Veligura$^1$, H.T. Jonkman$^2$}
\author{B.J. van Wees$^1$}

\affiliation{$^1$Physics of Nanodevices, Zernike Institute for Advanced Materials, University
of Groningen, Nijenborgh 4, 9747 AG Groningen, The Netherlands\\
$^2$Molecular Electronics, Zernike Institute for Advanced Materials, University
of Groningen, Nijenborgh 4, 9747 AG Groningen, The Netherlands
}

\date{\today}

\begin{abstract}

Spin transport in graphene carries the potential of a long spin diffusion length at room temperature. However, extrinsic relaxation processes limit the current experimental values to 1-2 $\mu$m. We present Hanle spin precession measurements in gated lateral spin valve devices in the low to high (up to 10$^{13}$ cm$^{-2}$) carrier density range of graphene. A linear scaling between the spin diffusion length and the diffusion coefficient is observed. We measure nearly identical spin- and charge diffusion coefficients indicating that electron-electron interactions are relatively weak and transport is limited by impurity potential scattering. When extrapolated to the maximum carrier mobilities of 2$\times 10^5$ cm$^2$/Vs, our results predict that a considerable increase in the spin diffusion length should be possible.

\end{abstract}

\maketitle

The high charge carrier mobility \cite{HighMob,HighMob2,ballisticgraphene} and spin diffusion length of micrometers \cite{Tombros2007,Kawakami2008} measured at room temperature make graphene a possible candidate for future electronic and spintronic devices. This two-dimensional crystalline material has two electronic conduction regimes, metallic where charge carriers are of one type (holes or electrons), and the region around the Dirac neutrality point where transport of electric current happens through small regions charged locally with holes or electrons ("electron-hole puddles"). The presence of such puddles yielding a finite local density $|n| \approx 10^{11}$ cm$^{-2}$ was shown experimentally using scanning single electron transistor technique\cite{puddlesSET} and scanning electron spectroscopy\cite{puddlesSTS1,puddlesSTS2}, with the intensity of the fluctuations being strongly enhanced by substrate impurities\cite{ballisticgraphene}.

Experiments done so far on spin transport reveal room temperature spin relaxation times of the order of 100 to 200 ps\cite{Tombros2007}. This is well below the theoretically predicted, intrinsic limit\cite{BrataasScattering,TheorSpinGraphene} but might be explained if we consider extrinsic effects\cite{CastroNeto}. Since hyperfine interactions at 300K are weak in graphitic systems, there are two possible mechanisms that can be held responsible for such a strong spin relaxation\cite{EYTheory}, scaling differently on the momentum relaxation. In case of the Elliot-Yafet mechanism (spin flip occurs with a finite probability at each momentum scattering center) the spin scattering time $\tau_s$ is proportional to $\tau_d$, the momentum scattering time, while the D'yakonov-Perel mechanism (spins precess under the influence of local spin-orbit fields in between scattering events) is characterized by $\tau_s \propto \tau_d^{-1}$. To identify the scattering mechanism and find the ultimate limit on spin relaxation, one can thus investigate the link between spin transport and the electronic quality of the graphene, in particular the charge carrier mobility. Since the mobility is ill-defined at or in the vicinity of the Dirac neutrality point, we will use the diffusion coefficient defined as $D=\frac{1}{2} v_F l$ and link it to the spin diffusion length $\lambda_s = \sqrt{D \tau_s}$. Here $v_F$ is the Fermi velocity and $l$ represents the scattering mean free path.

\begin{figure}[b!]
\includegraphics[width=8.5cm]{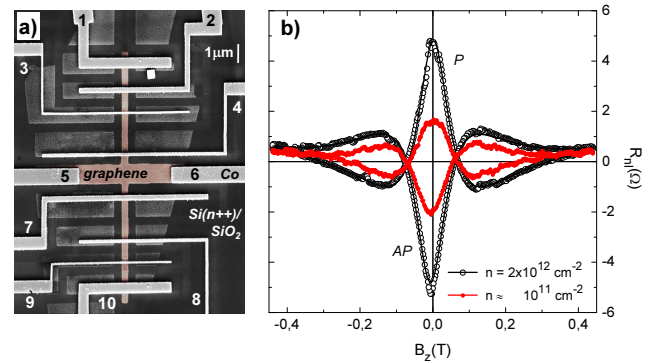}
\caption{\label{fig:device} (Color online.) a) SEM image of a spin valve device. A $0.3\times 13 \mu$m strip with a cross shape in the middle (brown/light gray) was etched with oxygen plasma out of the original graphene flake (dark gray). The Co electrodes 1-10 of widths 90 to 800 nm and spacings 1 to 3.1 $\mu$m are also visible. b) Hanle precession measurements (dots) and fits (solid lines) at the Dirac neutrality point and in the metallic regime, with the injector/detector magnetization aligned parallel(P) and antiparallel(AP).}
\end{figure}

One way to study charge against spin diffusion would be comparing the results in a set of devices that display significantly different carrier mobilities. However, it is experimentally challenging to fabricate consistently good ferromagnetic contacts to the graphene for such a set of samples. The option we choose here is to do the experiments on individual devices tuning the carrier density from the metallic regime down to the lowest values and comparing the behavior of the spin transport to the changes in the charge diffusion coefficient.

In this Rapid Communication we present a systematic study of the spin transport and scattering at room temperature in single layer graphene samples on SiO$_2$ substrate. The measurements are done at a wide range of carrier densities with an accent on the Dirac neutrality point where the transport is difficult to model and Coulomb electron-electron interactions are expected to be the strongest\cite{CoulombTheory}. We compare this directly to the charge transport in the same samples to learn more about the diffusion phenomena and the interactions that lead to spin relaxation.

The charge carrier transport in graphene in the metallic regime (at an energy $E$ sufficiently far away from the Dirac neutrality point) can be described by a diffusion process characterized by the 2-dimensional charge diffusion coefficient $D$. The density of states (DOS) is given by\cite{Ando2006}
\begin{equation}\label{Eq:DOS}
\nu(E)=\frac{g_v g_s 2 \pi |E|}{h^2v_F^2}
\end{equation}
with the twofold valley ($g_v=2$) and spin ($g_s=2$) degeneracies and the Fermi velocity $v_F\approx 10^6$ms$^{-1}$. By integration we can obtain the density $n(E_F)= g_v g_s \pi E_F^2 / (h^2v_F^2)$ with $E_F$ the Fermi energy, and the Einstein relation $\sigma = e^2 \nu D$ allows for calculating the charge diffusion coefficient
\begin{equation}\label{Eq:diffcoff}
D=\frac{1}{R_s e^2 \nu}=\frac{h v_F}{2e^2 \sqrt{g_v g_s \pi}}\frac{1}{R_s \sqrt{n}}.
\end{equation}
Here $e$ is the electron charge and $R_s$ is the square resistance of the graphene layer, inverse of the conductivity $\sigma$. Finally, using the semiclassical Drude formula one can calculate the carrier mobility $\mu=(R_s n e)^{-1}$ for the metallic regime.

In order to determine the charge and spin diffusion coefficients experimentally we have fabricated field effect devices where the single layer graphene flake is contacted by several ferromagnetic cobalt electrodes. A scanning electron microscope (SEM) image of such a device is shown in Fig.~\ref{fig:device}a). The graphene flakes are obtained by mechanical exfoliation from commercially available Kish graphite and deposited on a thermally oxidized, \textit{n++} doped Si substrate (300 nm thick oxide layer). The Si substrate contacted by a bottom Au electrode is used as electrostatic gate; applying a voltage $V_g$ of typically tens of volts on it allows reaching carrier densities in the graphene $|n| \leq 10^{13}$ cm$^{-2}$ (electrons or holes) calculated from the capacitance\cite{RiseofG}. A set of predefined Ti/Au markers help to accurately locate selected graphene flakes through optical- and atomic force microscope. The flakes are then etched with oxygen plasma into a cross shape, to allow for precise Hall type measurements using side contacts e.g. the ones labeled 5 and 6 in Fig.~\ref{fig:device}a) as current injectors and contacts 4 and 7 as Hall voltage probes. The electrical contacts are patterned using electron beam lithography and evaporated thermally at a base pressure of $\sim 10^{-6}$ mbar followed by standard lift-off technique. To achieve a high spin injection efficiency\cite{MisuPRB} a 0.8 nm thick Al$_2$O$_3$ insulating layer was introduced between the graphene and the ferromagnet, resulting in contact resistances of the order 20-40 k$\Omega$.

\begin{figure}[!t]
\includegraphics[width=8.5cm]{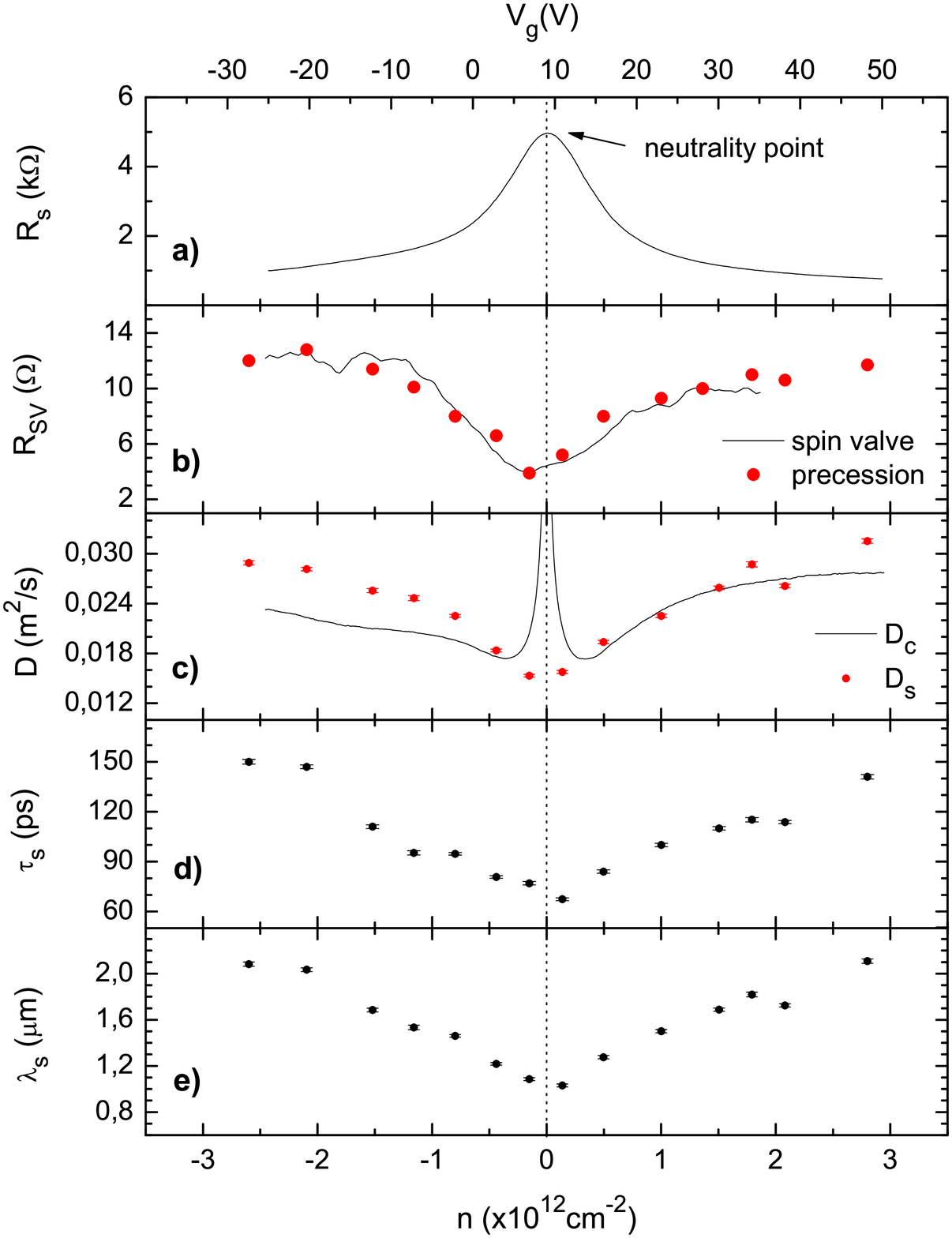}
\caption{\label{fig:meas} (Color online.) Charge- and spin transport parameters plotted against charge carrier density and gate voltage. a) Square resistance; b) Non-local spin valve signal determined from spin valve- and from Hanle precession measurements; c) Charge- and spin diffusion coefficients; d) Spin relaxation time; e) Spin relaxation length.}
\end{figure}

The graphene's square resistance $R_s$ is determined from four-probe local measurements. Sending an electric current from e.g. contact 1 to 4 and measuring the voltage drop between 2 and 3 we are sensitive only to the resistance of the graphene between contacts 2 and 3. Measuring the resistance against the applied gate voltage (i.e. in function of carrier density) and normalizing it to the graphene length to width ratio yields the $R_s$ curve plotted in Fig.~\ref{fig:meas} a). From such measurements we calculate the charge diffusion coefficient $D_c$ in the metallic regime using relation~\ref{Eq:diffcoff}, see the solid line in Fig.~\ref{fig:meas}c). The decrease in carrier density comes with a decrease in the diffusion coefficient; the singularities in the calculated $D_c$ at the charge neutrality point will be discussed later. The asymmetry in the electron- versus hole diffusion at high densities visible in panel c) probably originates from nonuniformities in the carrier density and can be traced back to the $R_s$ measurement in panel a). Measurements of the Hall coefficient $R_H$ (not shown) against the gate voltage using the cross contact geometry indicate the onset of the metallic regime at a carrier density $n \simeq \pm 0.5 \times 10^{12}$ cm$^{-2}$ by displaying a clear $1/n$ dependence. The density value extracted from the Hall measurements in the metallic regime confirms the number calculated from the square resistance measurements and gate capacitance.

The spin transport measurements are performed in the non-local geometry\cite{Tombros2007}: a spin-polarized current is injected e.g. through electrode 2 and extracted through electrode 1, while we measure the voltage between electrodes 3 and 4. There is no charge current flowing between 3 and 4; the detected non-local signal $R_{nl}$ in an in-plane magnetic field is purely due to the effect of spins diffusing from the injector electrodes to the detectors. Subtracting $R_{nl}$ at parallel and antiparallel magnetic orientation of the injector/detector electrodes while scanning the gate voltage gives the spin valve signal $R_{SV}$ that has a significant dependence on the charge carrier density as plotted in Fig.~\ref{fig:meas} b), solid line. The electrodes 1 and 4 are far enough not to contribute significantly to $R_{SV}$, therefore we define $L=3.1 \mu$m as the distance between the two inner electrodes.

Applying a magnetic field $B_z$ orthogonal to the sample plane will result in Hanle spin precession. Measuring $R_{nl}$ while we sweep the magnetic field (i.e. we change the precession frequency) yields the curves in fig.~\ref{fig:device}b). Here two measurements are plotted for the metallic regime, $V_g=+40V$ ($n \simeq -2 \times 10^{12}$ cm$^{-2}$), and two for the Dirac neutrality point, $V_g=+9V$, with the central injector/detector electrodes oriented parallel and antiparallel, respectively. The parallel-antiparallel signal difference at zero field is the same as the spin valve signal defined above and is plotted for different densities  in Fig.~\ref{fig:meas}b), dots. The advantage of a spin precession measurement is that it allows extracting the spin diffusion coefficient $D_s$ and spin scattering time $\tau_s$ by fitting the measurements with the solutions to the Bloch equation\cite{EYTheory} for spin accumulation $\mathbf{\mu}$:
\begin{equation}\label{Eq:Bloch}
D_s \nabla^2 \mathbf{\mu} - \frac{\mathbf{\mu}}{\tau_s} + \frac{g \mu_B}{\hbar} \mathbf{B} \times \mathbf{\mu}=0
\end{equation}
where the first term on the left hand side describes the spin diffusion, the second term the spin relaxation and the last one the precession, with an effective Land\'{e} factor $g=2$ and the Bohr magneton $\mu_B$.

A set of precession measurements was done for different charge carrier densities; the resulting spin transport parameters $D_s$ and $\tau_s$ are plotted in Fig.~\ref{fig:meas}c) and d). The spin diffusion length $\lambda_s = \sqrt{D_s \tau_s}$ is shown in panel e).

Examining Fig. 2, we see that $D_s$, $\tau_s$ and $\lambda_s$ all decrease approximately by a factor of 2 when we approach the neutrality point. This results in a strong decrease in the detected spin valve signal as seen in Fig.~\ref{fig:meas}b), consistent with the prediction of the formula for a four-terminal non-local spin injection geometry\cite{Jedema}:
\begin{equation}\label{Eq:Jedema}
R_{SV} = \frac{P^2 R_s \lambda_{s}}{W_g} \exp(-L/\lambda_{s}),
\end{equation}
where $W_g = 300$nm is the width of the graphene flake. The spin polarization of the injected current determined from this relation is $P\simeq 9\%$. Note that this value can be considered constant through the range of carrier densities we used, since the contact resistances span from 20 to 40~k$\Omega$ where impedance mismatch is suppressed\footnote{We have considered the spin injection model for transparent contacts from \cite{MisuPRB} that corrects formula~\ref{Eq:Jedema} addressing the impedance mismatch. The influence of the correction on the injection efficiency is below 10\%.}.

Let us focus now on the diffusion of charge versus spin. As visible in Fig.~\ref{fig:meas}c), in the high density case the values are practically identical for spin and charge. This is a striking observation, since the two physical entities, $D_c$ and $D_s$, are determined from completely different types of experiments.

\begin{figure}
\includegraphics[width=8.5cm]{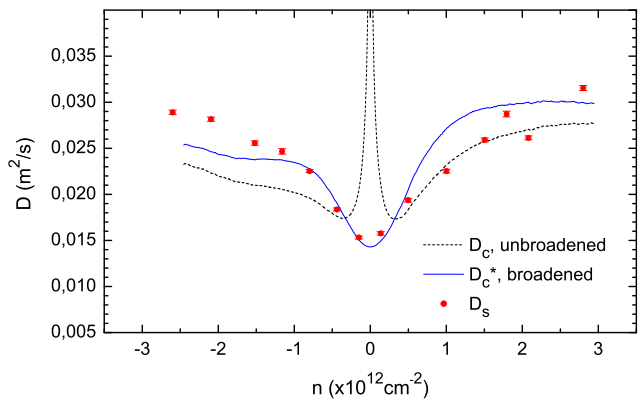}
\caption{\label{fig:DcvsDs} (Color online.) Spin- vs. charge diffusion coefficient with the unbroadened DOS from Eq.~\ref{Eq:DOS} and the broadened version from Eq.~\ref{Eq:BroadenedDOS} using a Gaussian broadening of FWHM$\simeq 176$ meV.}
\end{figure}

However, for $|n|<0.5 \times 10^{12}$cm$^{-2}$ formula~\ref{Eq:DOS} yields unphysical values for the diffusion coefficient and results in a singularity at the Dirac neutrality point. This comes from the unrealistic assumption of vanishing carrier density and DOS. To correct for it, one has to account for a broadened density of states $\nu^*(E)$ due to finite temperature, electron-hole puddles and possibly to the finite lifetime of electronic states. The simplest way to include all broadening effects in the DOS is to add a Gaussian broadening energy $\sigma$ in the form of
\begin{equation}\label{Eq:BroadenedDOS}
\nu^*(E)=\frac{1}{\sqrt{2\pi}\sigma} \int_{-\infty}^{\infty} \exp\left(-\frac{(\epsilon-E)^2}{2 \sigma^2}\right) \nu(\epsilon) d\epsilon.
\end{equation}
After integration we obtain
\begin{equation}\label{Eq:FinalDOS}
\nonumber \nu^*(E)=\frac{g_v g_s 2 \pi}{h^2 v_F^2} \left[ \frac{2 \sigma}{\sqrt{2 \pi}} \exp \left(-\frac{E^2}{2\sigma^2} \right) + E ~\mbox{erf} \left(\frac{E}{\sigma \sqrt{2}} \right) \right]
\end{equation}
where $\mbox{erf}$ is the Gaussian error function and the only undetermined parameter is the value of $\sigma$.
Replacing the DOS with the broadened version in formula~\ref{Eq:diffcoff} we  plot the modified diffusion constant $D_c^*$ in function of the density together with the unmodified charge- and spin diffusion constants, see Fig.~\ref{fig:DcvsDs}. We find good correspondence between $D_c^*$ and $D_s$ both at low and high densities if and only if we choose an energy broadening of $\sigma \simeq 75$meV~\footnote{The value of $\sigma$ is chosen to cure the singularity at $n \approx 0$ by fitting the broadened $D_c^*$ to $D_s$ for densities $|n|< 0.5 \times 10^{12}$ cm$^{-2}$. At $|n|\gtrsim3 \times 10^{12}$ cm$^{-2}$ the broadened curve approaches the unbroadened one determined from the measurements.}, i.e. a Gaussian with full width at half maximum (FWHM) = $2\sqrt{2ln2}\sigma \simeq 176$ meV corresponding to a density variation of $\Delta n \simeq \pm 0.7 \times 10^{12}$ cm$^{-2}$. This is consistent with the literature values  \cite{puddlesSTS1} attributed to electron-hole puddles in graphene on SiO$_2$, considering that our samples show a carrier mobility of only 3000 cm$^2$/Vs.\cite{JozsaDrift}

The observation that the spin diffusion coefficient shows no considerable difference from the charge diffusion coefficient indicates a minor role of Coulomb electron-electron interactions\cite{SpinDrag}. This is in agreement with the recent results of Li et al.\cite{eeinteract} where electron-electron interactions are detected in high carrier mobility, suspended graphene flakes only. The major mechanism for limiting the (spin) diffusion seems to be the impurity potential scattering.

\begin{figure}[t!]
\includegraphics[width=8.5cm]{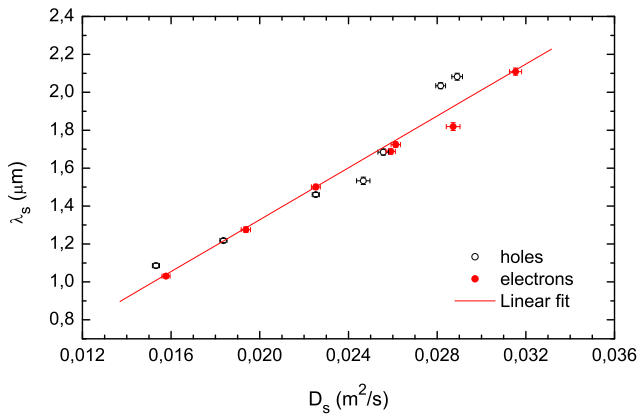}
\caption{\label{fig:DvsLambda} (Color online.) Linear relationship between the spin relaxation length and the spin diffusion coefficient, extracted from Fig. 2 panels c) and e).}
\end{figure}

The most convincing argument comes, however, from the scaling between (spin- or charge) diffusion coefficient and spin diffusion length. Plotting the values of $\lambda_s$ against $D_s$ as extracted from Fig.~\ref{fig:meas} shows a clear linear dependence for both the electron and the hole conduction regime, see Fig.~\ref{fig:DvsLambda}. Since $\lambda_s = \sqrt{D \tau_s}$, the linear dependence means that the spin scattering time is directly proportional to the diffusion coefficient, i.e. to the momentum scattering time. The experiments confirm thus an Elliot-Yafet type spin relaxation mechanism, in agreement with our earlier spin relaxation anisotropy studies presented in Ref.~\onlinecite{TombrosAnisotropy}.

In conclusion, we expect that improving the electronic characteristics of the graphene flake by e.g. removing the substrate (suspended graphene), annealing with high electric currents and/or using selected starting material (higher purity graphite) shall both enhance the charge transport and prolong the spin scattering time. Assuming the Elliot-Yafet mechanism is still dominating at high carrier mobilities we can extrapolate the behavior shown in Fig.~\ref{fig:DvsLambda} to samples displaying charge carrier mobilities in the range of $2 \times 10^5$ cm$^2$/Vs reported recently, to predict a possible room temperature spin diffusion length up to 100 micrometers~\footnote{In metals it is known that the ratio between spin- and momentum relaxations is approximately the same for scattering induced by impurities or electron-phonon interactions. We assume that this is also the case in graphene.}.

This work is part of the research programme of the Foundation for Fundamental Research on Matter (FOM), which is financially supported by the Netherlands Organisation for Scientific Research (NWO). We acknowledge the financial support of the Zernike Institute for Advanced Materials and NanoNed.

\end{document}